\documentclass[fleqn,usenatbib]{mnras}

\usepackage{bbm}
\usepackage{mathrsfs}
\usepackage{amssymb,amsmath,float}
\usepackage{rotating}
\usepackage{color}

\newcommand\cE{\mathcal{E}}

\newcommand\de{\delta}
\newcommand\ep{\epsilon}

\renewcommand\th{\theta}

\newcommand\ka{\kappa}
\newcommand\lam{\lambda}

\newcommand\si{\sigma}
\newcommand\ta{\tau}

\newcommand\om{\omega}

\newcommand\De{\Delta}

\newcommand\ie{\emph{i.e.}}
\newcommand\eg{\emph{e.g.}}

\newcommand\beq{\begin{equation}}
\newcommand\eeq{\end{equation}}
\newcommand\bea{\begin{eqnarray}}
\newcommand\eea{\end{eqnarray}}
\newcommand\bal{\begin{align}}
\newcommand\eal{\end{align}}

\newcommand\fr{\frac}

\newcommand\ap{\approx}

\renewcommand\d{\mathrm{d}}

\newcommand\bj{\bold{j}}
\newcommand\bk{\mathbf{k}}

\newcommand\bu{\mathbf{u}}

\newcommand\bR{\bold{R}}

\renewcommand\bal{\mbox{\boldmath$\alpha$}}

\title[EUV]{A model of EUV emission from clusters of galaxies}

\author[R. Lieu \& C.-H. Shi]{Richard Lieu$^{1}$\thanks{E-mail:lieur@uah.edu}
Chun-Hui Shi$^{2}$
\\
$^{1}$Department of Physics and Astronomy, University of Alabama, Huntsville, AL 35899\\
$^{2}$Chinese Academy of Sciences South America Center for Astronomy,\\
\quad China-Chile Joint Center for Astronomy (CASSACA), Camino El Observatorio  1515, Las Condes, Santiago, Chile
}

\date{Accepted XXX. Received YYY; in original form ZZZ}

\pubyear{2022}

\begin{document}
\label{firstpage}
\pagerange{\pageref{firstpage}--\pageref{lastpage}}
\maketitle

\begin{abstract}
With tantalizing evidence of the recent e-Rosita mission, 
re-discovering very soft X-rays and EUV radiation from a cluster of galaxies or its environment, 
the question of the origin of cluster EUV excess is revisited in this work.  
It will be shown that the gas temperature, 
density, and frozen-in magnetic field of the intracluster medium, collectively support the emission and propagation of coherent \u{C}erenkov radiation, 
which is low frequency and large amplitude radiation capable of accelerating charged particles to relativistic speeds.  
Owing to the spectrum of \u{C}erenkov radiation, most of the incipient relativistic electrons undergo inverse-Compton scattering with the cosmic microwave background.  It turns out the scattered radiation has observable ramifications only in the EUV band, of photon energy $70 -- 100$~eV, 
having a luminosity $\ap 10^{44}$~ergs~s$^{-1}$.  This luminosity is on par with the EUV excess level detected from Abell 1795 and the Coma cluster.  It should be stressed, as {\it caveat emptor}, that although the main subject is the putative large amplitude coherent \u{C}erenkov modes which are highly nonlinear, the results presented were derived using a quasi-linear approach to highlight the observable features of the phenomenon, namely the EUV emission.
\end{abstract}




\section{Introduction}

With the advent of the e-Rosita mission,
there has recently been renewed interest in the phenomenon of EUV excess in clusters discovered by \citet{lie96,lieu96,mit98}, 
because first-light observations revealed preliminary evidence 
in support of spatially localised $0.55 - 0.65$ $keV$ emissions 
in the outskirts of the Coma cluster and between the A3391/95 cluster pair \citep{chu21,rei21}.  
Although this energy range corresponds to the OVII and OVIII emission lines, 
both symptomatic of warm ($\ap 10^6$ K gas), 
no detailed spectroscopic study has yet been done to demonstrate their existence unequivocally at the redshift of the clusters. 
Such a study is currently ongoing, and it is essential because of a long-held view \citep{hwa97,ens98,sar98} that the EUV excess could also be due to inverse-Compton (IC) interaction between a population of relativistic electrons and the cosmic microwave background (CMB).

The purpose of this paper is to develop the theme of a non-thermal origin of the EUV excess further, to show that the observational revelation of the intracluster medium (ICM) carrying a frozen-in magnetic field enables the free electrons in the ICM plasma to emit coherent \u{C}erenkov radiation. 
This radiation consists of large amplitude and low-frequency electromagnetic waves (helicon waves) propagating through the ICM despite their frequencies lying below the plasma frequency.  
The waves are subsequently either reabsorbed by the ICM or dissipated via IC scattering between the CMB and relativistic electrons accelerated by the waves. 
The output of the latter is EUV emission of brightness on par with the measured EUV excess.

\section{Characteristics of the magnetoionic ICM; refractive index}
\label{sec:characteristics}

The emission by and propagation of low-frequency radiation in the ICM are two aspects of the physics of clusters of galaxies which have not yet been very well investigated.  
To the lowest order of approximation, the ICM may be considered as a $kT \approx$ 10~keV,~$n_e \approx 10^{-3}$~cm$^{-3}$ 
plasma filling a uniform sphere of radius $\approx$~1~Mpc \citep{mir20} smoothly, 
and having a frozen-in magnetic field $B \approx 1~\mu$G, 
which is coherent over the scale $\ell_c \approx 1$~Mpc 
\citep[more precisely $B \approx 2-6 (\ell_c/10~{\rm kpc})^{-1/2}~\mu$~G, see][]{boh16}.  
The plasma frequency is given by 
\beq \om_p = 1.78 \times 10^3 \left(\fr{n_e}{10^{-3}~{\rm cm}^{-3}}\right)^{1/2}~{\rm rad~s}^{-1}, 
\label{omp2} 
\eeq 
and the electron cyclotron frequency is
\beq 
\om_b
= \fr{eB}{m_ec}
=17.6\left(\fr{B}{1~\mu G}\right)~{\rm rad\ s^{-1}} \label{cyclotron}\eeq
Thus  $\om_b \ll \om_p$ from \eqref{omp2} and \eqref{cyclotron}.

Unlike an unmagnetised plasma, radiation in a very low frequency,
which satisfies the inequality 
\beq 
\om \ll \om_b \ll \om_p, \label{plasmacond}
\eeq
and has a prescribed polarization, 
can propagate through a plasma with a frozen-in magnetic field.
Specifically, if one defines the three dimensionless parameters
\beq 
\xi=\fr{\om_p^2}{\om^2},\ \eta=\fr{\om_b^2}{\om^2}, \label{xi_eta}
\eeq 
and the angle $\theta$
between the magnetic field and the radiation propagation direction,
the refractive index $n$ of the medium as given by 
\citet{jin06}
simplifies, in the regime $\xi \gg \eta \gg 1$,
to assume the form
\beq 
n^2=1-\xi \left[1 +\fr{\eta\sin^2 \th}{2\xi} \pm \left( \fr{\eta^2\sin^4\theta}{4\xi^2}+\eta{\rm cos}^2 \theta \right)^{1/2}\right]^{-1}.\label{etasq}
\eeq

Strictly speaking, there is a contribution from the ions in the form of an additional term on the right side of (\ref{etasq}),
which is negligible 
for $\om_b \gtrsim\om \gg \om_{\rm bi} = ZeB/m_i$, 
but, for $\om \ll \om_{\rm bi}$, 
is identical to the $-\xi [\cdots]^{-1}$ part of \eqref{etasq} except with the substitutions $\xi\to\xi_i = \om_{\rm pi}^2/\om^2 $, $\eta\to\eta_i = \om_{\rm bi}^2/\om^2 $,  and $\pm\to\mp$, where $\om_{\rm pi}$ is the ion plasma frequency and  $\om_{\rm bi}$ is the ion cyclotron frequency.  
For a pure hydrogen plasma $\om_{\rm pi}$ and $\om_{\rm bi}$ are readily evaluated by the replacement $m_e \to m_p$ in the original expressions for $\om_p$ and $\om_b$ (namely $\om_p \propto m_e^{-1/2}$ and $\om_b \propto m_e^{-1}$).  
In the regime, $\om\approx\om_{\rm bi}$, the expression for $n^2$ is more complicated than \eqref{etasq} and the aforementioned extra term, which, nevertheless, is less relevant because it applies to a very narrow spectral range, and because the EUV emission from clusters is due ultimately to large amplitude \u{C}erenkov radiation in the $\om_b \gtrsim\om \gg \om_{\rm bi}$ regime (see below) where \eqref{etasq} is sufficiently accurate.

For plasma conditions satisfying \eqref{plasmacond}, we have 
\beq 
\xi \gg \eta \gg 1,\label{xi_eta_relation}
\eeq
and \eqref{etasq} reveals that apart from a narrow range of angles within the interval 
\beq 
\delta \theta \lesssim \fr{\sqrt{\eta}}{2\xi} \leq 5 \times 10^{-5} \label{cone} 
\eeq 
on either side of $\theta=\pi/2$, 
the $\cos^2\theta$ term in \eqref{etasq} dominates the $\sin^4\theta$ and $\sin^2\theta$ terms,
~\ie\, for the vast majority of propagation directions, one may ignore the $\sin^4\theta$ and $\sin^2\theta$ terms.  
In this way, \eqref{etasq} simplifies to 
\beq 
n^2=1-\fr{\om_p^2/\om^2}{1\pm\fr{\om_b}{\om}{\rm cos}\theta};~\om \ll \om_b \ll \om_p,~ \label{etasq_2}
\eeq
where the $+$ and $-$ signs denote the left and the right-handed polarization of the radiation, respectively.  
For ultra-low frequencies $\om \ll \om_{\rm bi}$ then, in the case of a hydrogen plasma,  
\beq
\begin{split}
n^2=  1-\fr{\om_p^2/\om^2}{1\pm\fr{\om_b}{\om}{\rm cos}\theta} -
& \fr{\om_{\rm pi}^2/\om^2}{1\mp\fr{\om_{\rm bi}}{\om}{\rm cos}\theta};\\
& \qquad
 \om \ll \om_{\rm bi}~{\rm and}~\om_b \ll \om_p. \label{etasq_3}
\end{split}
\eeq

For radiation frequency satisfying
\eqref{plasmacond} with the $-$ mode propagating along directions 
$\cos\th >0$ and the $+$ mode along $\cos\th < 0$, 
(again, $\th$ well avoiding the very narrow `forbidden cone' defined by \eqref{cone}), 
$n^2$ is $\gg 1$, and one may access the salient features of the propagation by setting $\cos\th = 1$ to simplify \eqref{etasq_2} and \eqref{etasq_3} to become, 
in the case of a hydrogen plasma 
\beq
n \ap \fr{\om_p}{\sqrt{\om\om_b}} \Theta(\om - \om_{\rm bi}) 
+ \fr{\om_p}{\sqrt{\om_{\rm bi} \om_b}} \Theta (\om_{\rm bi} - \om) \label{nrni} 
\eeq
where $\Theta (\om)$ is the Heaviside unit step function.  
The derivation of this result is shown in Appendix \ref{appendix::RefractiveIndex}.
Thus the wave vector of the radiation is  
\beq
k = \fr{n_r\om}{c} \ap \fr{\om_p}{c} \left(\fr{\om}{\om_b}\right)^{1/2},~\om\ll\om_b\ll\om_p. \label{krki} 
\eeq    
Note that the aforementioned $1$~Mpc coherence length of the $1~\mu$G field assures the invariance of the radiation propagation direction and polarization 
(both being defined with reference to the magnetic field orientation).

It is also important to point out that for the modes of radiation emitted outside the forbidden cone \eqref{cone}, 
they can propagate because plasma electrons are predominantly affected by them in directions along the frozen-in magnetic field.  
Thus the frozen-in field is not affected by such modes, but rather by the modes emitted {\it inside} the cone.  
But since the total energy density of \u{C}erenkov radiation $U_\textrm{\u{C}erenkov}$ cannot exceed the hot ICM, \eqref{EU}, 
and \u{C}erenkov radiation energy density within the cone is only $\ap 10^{-5}U_\textrm{\u{C}erenkov}$ even when assuming emission and propagation are possible, 
this is far less than the energy density of the frozen-in field, 
which at a few $\mu$~G field strength is $\ap$ a few \% of the hot ICM.  
Thus the overall conclusion is that \u{C}erenkov radiation does not exert any disruptively large perturbations to the model of an ICM with a frozen- field 
at strengths commensurate with observed values.

Evidently, based on the requirement given after \eqref{etasq_2}, the magnetoionic ICM is transparent to $\ka\approx 50\%$ of unpolarized radiation emitted by an isotropic source at frequencies $\om \lesssim \om_b$. 
With these in mind, we shall henceforth assume that the $n$ is given by the simplified version of \eqref{etasq_2}, \ie~\eqref{nrni}.

\section{Coherent \u{C}erenkov radiation and its reabsorption in the magnetoionic ICM}
\label{sec:coherent}

Having demonstrated in section \ref{sec:characteristics} that the possibility of a large and positive refractive index for the ICM, 
we turn to the problem of emission by an ensemble of electrons,
namely the coherent emission from a volume of size one wavelength or less.
We first consider a single charge $e$ moving at velocity $\mathbf{u}$ with a component $\mathbf{u}_\parallel\,\parallel\mathbf{B}$, $(u_\parallel\ll c)$
due to thermal agitation
(motion transverse to $\mathbf{B}$ is not uniform
and gives rise to cyclotron radiation which is not the topic of this paper).
Although this derivation is textbook material, 
we are still going through it with some care to pave the way to treat coherent \u{C}erenkov emissions. 
The position and velocity of the charge are given by
\begin{equation}
\mathbf{r} = \mathbf{u}_\parallel t;
\qquad \mathbf{v} = \mathbf{u}_\parallel
\end{equation}
The ensuing electric current density is
\begin{equation}
\mathbf{j}(\omega,\,\mathbf{k}) = 
e\int_{-\infty}^{+\infty}dt ~
\mathbf{v}(t)\,
e^{i(\omega t-\mathbf{k}\cdot\mathbf{r})}
= 2\pi e \mathbf{u}_\parallel\delta(\mathbf{k}\cdot\mathbf{u}_\parallel -\omega)
\end{equation}
The next step towards the radiation emission rate is to evaluate $\mathbf{k}\times\mathbf{j}$, as
\begin{equation}
\mathbf{k}\times\mathbf{j} = 2\pi e (\mathbf{k}\times\mathbf{u}_\parallel)
\delta(\mathbf{k}\cdot\mathbf{u}_\parallel-\omega).
\label{SingleChargeCross}
\end{equation}

The total emission energy (in Joules) is given by 
\begin{equation}
\mathcal{E} = \frac{1}{2\pi^3}\int_0^\infty\omega d\omega
\int \frac{d^3\mathbf{k}}{k^2}\,\textrm{Im}\bigg(\frac{|\mathbf{k}\times\mathbf{j}|^2}{k^2c^2-\epsilon\omega^2}\bigg),
\label{EmissionEnergy}
\end{equation}
where $\textrm{Im}(\cdots)$ denotes the imaginary part, 
and $\epsilon(\omega,~\mathbf{k})=\epsilon_1(\omega_1,~\mathbf{k}) + i\epsilon_2(\omega_2,~\mathbf{k})=n^2$
is the dielectric constant of the medium into which the emission takes place.
Substituting \eqref{SingleChargeCross} into \eqref{EmissionEnergy},
one finds a spectral emission rate,
summed over all the directions of $\mathbf{k}$, of 
\begin{equation}
\begin{split}
\frac{d^2\mathcal{E}}{dtd\omega} = \frac{e^2\omega}{\pi^2} & 
\int_0^{2\pi}d\phi\int_0^\pi\sin^2\theta d\cos\theta\\
\quad
 & \times\int_0^\infty k^2dk\,
\textrm{Im}\bigg[\frac{u^2_\parallel\delta(\mathbf{k}\cdot\mathbf{u}_\parallel-\omega)}{k^2c^2-\epsilon\omega^2}\bigg],
\label{SingleEmissionRate}
\end{split}
\end{equation}
where the relation $\delta^2(\omega)=T/(2\pi)$ with $T$ being the 
'total duration of emission' was employed.
It is also assumed that $\theta$ and $\phi$ are the polar and azimuthal angles that $\mathbf{k}$
makes with $\mathbf{u}_\parallel\parallel\hat{\mathbf{z}}$
({\it i.e.} $\hat{\mathbf{z}}$ marks the direction of the frozen-in magnetic field $\mathbf{B}$).

After performing the $k$ integration in \eqref{SingleEmissionRate}
\bea 
\fr{d^2\mathcal{E}}{dt d\om} &=& \fr{q^2\om}{\pi c^2} \int^1_{-1} u_\parallel \tan^2 \th~d\cos^2\theta~{\rm Im} \left(\fr{1}{1-\ep\beta^2{\rm cos}^2\theta}\right) \notag \\
&=& \fr{q^2 \om}{c^2} \int_{-1}^1 \sin^2 \th~d\cos^2 \th~\fr{u_\parallel}{\pi (\cos^2 \th - \cos^2 \th_0) + \Gamma^2} \notag\\
&=& \fr{q^2 \om u_\parallel}{c^2} \int_{-1}^1 \sin^2 \th~d\cos^2 \th~\de (\cos^2 \th - \cos^2 \th_0), \notag\\
\label{SingleEmissionRate2}
\eea
where ${\rm cos}^2\theta_0=\ep_1/(\beta^2\vert\ep\vert^2)$ 
and $\Gamma\equiv\epsilon_2/(\beta^2\vert\ep\vert^2)$, 
The $\delta$ functions are the limiting case of no absorption $\ep_2 \rightarrow 0$ (hence $\Gamma \rightarrow 0$, which is the assumption we are making about the IGM).

Integration of \eqref{SingleEmissionRate2} w.r.t. $\om$ yields  
\beq 
\fr{d\mathcal{E}}{dt} = \fr{q^2 u_\parallel}{c^2} \int_0^\infty \om \left(1- \fr{1}{n^2 \beta^2} \right) d\om 
\label{TotalEmissionRate}
\eeq 
which is the total emission rate of \u{C}erenkov radiation, a standard result. 
Note that the upper limit of integration in \eqref{TotalEmissionRate} 
is determined by the condition $n (\om) \beta =1$.  
From \eqref{etasq_2}, \eqref{nrni} and the discussion in between, 
one sees that while $n\gg 1$ for $\om \ll \om_b$, 
it approaches unity as $\om \to \om_b$ from below.  
Thus the integral terminates at $\om \approx \om_b$.   
Moreover, since $d\cE/dt \propto u_\parallel$, 
and under the assumption of a thermalized ICM 
the proton thermal velocity is much less than the electron, 
we shall ignore the emission of protons.

Turning to the problem of emission by an ensemble of electrons, attention is drawn to the possibility of coherent emission from volumes of size one wavelength or less.  
To see this, one writes the total current due to $N$ proximate electrons as 
\beq 
\mathbf{j} (\om,\bk) = \sum_{l=1}^N \int \bu^\parallel_l~e^{i[\om t - \bk\cdot (\bu^\parallel_l t + \bR_l)]}. 
\label{TotalCurrent}
\eeq  
In \eqref{TotalCurrent}, attention is drawn to the phase term $i\bk\cdot\bR_1$ in the exponent, 
which takes account of the effect of the various electrons being in different positions.  
For constructive interference among the terms within the $l$-summation to occur when one evaluates $|\bk \times \bj|^2$, 
the phases must differ from each other by an amount $\ll 2\pi$ within each set of $N$ electrons. 
Since $k=n\om/c$, this requires the position of the charges to occupy a volume $\lesssim (2\pi/k)^3=8\pi^3c^3/(n^3\om^3)$.  
The maximum value of $N$ comprising the largest coherence volume is then 
\beq N(\om) =\fr{8\pi^3c^3 n_e}{n^3\om^3} 
\label{N} \eeq 
where $n_e$ is the electron number density. Thus $N=N(\om)$ is a function of the radiation frequency $\om$.

Provided $N$ stays within the maximum allowed value for coherence emission, 
the phase $i\bk\cdot\bR_l$ defaults to $i\bk\cdot\bR_0$ 
where $\bR_0$ is constant among the $N$ electrons. 
Thus 
\beq 
\begin{split}
|\bk \times \bj|^2 = 
8\pi^2 q^2 \sum_{\xi \neq \eta}^{N(\om)}&  \sum_\eta^{N(\om)}  
(\bk \times \bu^\parallel_\xi) 
\cdot  (\bk \times \bu^\parallel_\eta) \\
& \!\!\times\de (\om - \bk\cdot\bu^\parallel_\xi)\cdot \de (\om-\bk\cdot\bu^\parallel_\eta),
\label{kxjsq} 
\end{split}
\eeq  
where $\xi$ and $\eta$ are summation indices, distinguishing from those dimensionless parameters in the last section.

Substituting into \eqref{EmissionEnergy}, 
the integral over $k$ becomes
\beq
\begin{split}
 &\int dk ~\textrm{Im}\bigg[\fr{\delta(\om-\textbf{k}\cdot\textbf{u}^\parallel_\xi)\delta(\om-\textbf{k}\cdot\textbf{u}^\parallel_\eta)}
{(1-\ep\om^2/c^2k^2)} \bigg]\\
= & \fr{1}{u^\parallel_\xi \cos\theta_\xi}
\textrm{Im}\left(\fr{1}{1-\ep\beta_\xi^2 \cos^2\theta_\xi}\right)\cdot
\delta\left(\fr{u^\parallel_\eta \cos\theta_\eta}{u^\parallel_\xi \cos\theta_\xi} \om-\om\right)
\label{c19}
\end{split}
\eeq
The $\de$-function is survived (\ie~it becomes $T/2\pi$) by a fraction $\ap 1/\pi$ of the $N(\om)$ electrons in the $\eta$-summation of \eqref{kxjsq}, 
for which $\th_\eta = \th_\xi$ but $\phi_\eta$ is independent of $\phi_\xi$ 
(we assumed for simplicity that the magnitude (speed) $u_\eta = u_\xi$ for all $\eta,\,\xi$).

In this way, after summing over an isotropic distribution of $\bk$ directions,  one obtains
\begin{equation}
\begin{split}
\fr{d^2 \cE}{dt d\om} =& \fr{2e^2 \om u_\parallel}{\pi c^2} N^2 (\om) \\
                       & \qquad\times\int_{-1}^1\sin^2 \th d\cos^2 \th~\de (\cos^2 \th - \cos^2 \th_0) \\ 
                      =& \fr{2e^2 \om u_\parallel}{\pi c^2} N^2 (\om) \left(1-\fr{1}{n^2 \beta^2}\right). \label{d2Edtdw}
\end{split}
\end{equation}
As the spectral emissivity of one coherence volume (of size $\lesssim$ one wavelength at the frequency $\om$), in units of ergs~s$^{-1}$~Hz$^{-1}$, and in the {\it far field} limit.  Note also that the $\mp$ circular polarization modes for the $\cos\th ^>_< 0$ hemispheres, namely the propagating modes with positive $n^2$ (see the discussion after (\ref{etasq_3})), reinforce each other when $\bk \times \bj$ in (\ref{kxjsq}) is projected onto the equatorial ($\th =0$) plane.

Next, we consider a very large region of the ICM of cosmological volume $V$ containing many contiguous coherent emission volumes.  
The total radiation power of this region is the sum of the power of each constituent coherence volume, \ie~
as setting $q$ to be $-e$ for electrons
\beq \fr{d^2 \cE}{d\om dt}\Bigg\vert_V  = \fr{V}{\lambda^3} \fr{d^2 \cE}{dt d\om} \propto \om^{-1/2}~{\rm for}~\om \leq \om_b, \label{differential} \eeq 
or
\beq
\begin{split}
\fr{d\cE}{dt}\Bigg\vert_V 
= & V\int^{\om_{\rm max}}_0\d\om ~ \fr{d^2\mathcal{E}}{dtd\om}\fr{1}{\lam^3}, 
\\
= & \fr{32\pi^2 e^2 u^\parallel c\om_b^{3/2} (\om_{\rm max}^{1/2} - \om_{\rm min}^{1/2}) n_e^2 \kappa V}{\om_p^3}. \label{lossform}    
\end{split}
\eeq
The formula applies to the $\om_{\rm bi} \leq \om \leq\om_b \ll\om_p$ regime, where $n$ is given by the first term on the right side of \eqref{etasq_3}.
Dividing by $V$, and applying \eqref{omp2}, \eqref{cyclotron}, and \eqref{nrni}, 
one obtains, for $\om_2 = \om_b$ and $\om_1 = \om_{\rm bi}$, 
\beq 
\begin{split}
\fr{1}{V}  \fr{d\cE}{dt}\Bigg\vert_V
& \!\!\!= 2.60 \times 10^{-10}  \left(\fr{n_e}{10^{-2}~{\rm cm}^{-3}}\right)^{1/2}\\
\!\!\!\times 
& \!\left(\fr{B}{1~\mu {\rm G}}\right)^2 
\left(\fr{kT}{10~{\rm keV}}\right)^{1/2}
\left(\fr{\kappa}{0.5}\right)~{\rm erg}~{\rm cm}^{-3}~{\rm s}^{-1}. \label{loss}    
\end{split}
\eeq
In \eqref{loss}, a parameter $\kappa$ is introduced to represent the fraction of emitted radiation with a polarization which enables it to propagate through the magnetoionic ICM at frequencies $\om \leq \om_b$, see the end of the last section.  For unpolarized radiation, $\kappa =0.5$.

Turning to the absorption coefficient, after taking into account stimulated emission, it is given by detailed balance in radiative transfer as 
\beq 
\begin{split}
\ka_{\rm abs} (\om) & = \fr{2\pi^2 c^2}{\hbar \om^3}(1-e^{-\hbar\om/kT}) \times \fr{1}{V}\fr{d\cE}{dt} \Bigg\vert_V \\
&\ap 1.76 \times 10^{15} \left(\fr{\om}{\om_b}\right)^{-5/2}~{\rm cm}^{-1} \label{abs}     
\end{split}\eeq
in the limit $\hbar\om \ll kT$ which certainly applies here.  
Since 
the mean free path $\ka^{-1}$ against absorption seems unphysically short, namely being less than the size of an atomic nucleus, before one concludes straightforwardly that this means at the relevant (and very low) frequencies $\om \lesssim \om_b$ the emission spectrum is a featureless black body, one should revisit the basics of radiation transport within a thermalized gas.

The key point to note is that, irrespective of the magnitude of $\ka$, when the absorption rate reaches the limit of \eqref{abs}, it has exceeded the stimulated emission rate fractionally by the amount 
\beq 
\eta = 1-e^{-\hbar\om/kT} \ap \frac{\hbar\om}{kT} \label{eta} 
\eeq 
expected of a Maxwellian thermal gas
\citep{bek66}
as such an excess would then exactly cancel the spontaneous emission rate of \eqref{loss} to ensure complete equilibrium.  Since $\ep\ap kT$, and $\eta \ap \De\ep/\ep$ where $\De\ep = \hbar\om$ is the energy difference between two quantum levels of the emitting electron, one sees that $\eta$ cannot be as stable as \eqref{eta} unless $\ep$ is the mean energy of a sufficiently large number $N$ of gas particles such that the relative fluctuation in the mean energy per particle as given by $\de\ep/\ep \ap 1/\sqrt{N}$ is at a level beneath $\De\om/\om$.  
This in turn sets a lower limit on the size $\ell$ of the black body radiator via 
\beq 
\begin{split}
\ell = n_e^{-1/3} \left( \fr{\hbar\om}{kT}\right)^{-2/3}
 & = 2.76 \times 10^{12} \left(\fr{n_e}{10^{-2}~{\rm cm}^{-3}}\right)^{-1/3}\\ 
&\times
\left(\fr{kT}{10~{\rm keV}}\right)^{2/3}
\left(\fr{\om}{\om_b}\right)^{-2/3}~{\rm cm}. \label{ell}     
\end{split}\eeq 
The true absorption length at each frequency is then given by \eqref{ell} rather than \eqref{abs}.  
The corresponding absorption time is 
\beq 
\begin{split}
\ta_{\rm abs} = \fr{n\ell}{c} 
= & 9.2 \times 10^3 
\left(\fr{n_e}{10^{-2}~{\rm cm}^{-3}}\right)^{1/6}~
\left(\fr{B}{1~\mu {\rm G}}\right)^{-1/2} \\
& \qquad\qquad\times
\left(\fr{kT}{10~{\rm keV}}\right)^{2/3}~
\left(\fr{\om}{\om_b}\right)^{-7/6}~s, \label{tabs}     
\end{split}
\eeq 
where $n$ is given by \eqref{nrni}.

\section{Synchrotron and inverse Compton emission by ICM electrons under the influence of \u{C}erenkov radiation }

With the behaviour of the ICM refractive index as described in the previous section in mind, 
it is shown in Section \ref{sec:coherent} that, for $\om < \om_b < \om_p$ the {\it propagating modes} of \u{C}erenkov radiation are emitted by the hot magnetoionic ICM as circularly polarized (right-handed in the forward hemisphere of the magnetic field, left-handed otherwise) having a large amplitude.  
Moreover, Section \ref{sec:coherent}
(especially the paragraphs embedding equations \eqref{TotalCurrent} - \eqref{d2Edtdw})
also derived the volume emissivity of coherent \u{C}erenkov radiation, {\it viz.}~radiation of wavelength $2\pi n\om/c > n_e^{-1/3}$, as 
\beq
\begin{split}
\fr{1}{V}  \fr{d\cE}{dt}\Bigg\vert_V
= & 2.46 \times 10^{-10} \left(\fr{n_e}{10^{-2}~{\rm cm}^{-3}}\right)^{1/2}
~\left(\fr{B}{1~\mu {\rm G}}\right)^2\\
& \!\!\!\times\left(\fr{kT}{10~{\rm keV}}\right)^{1/2}~\de\left(\fr{\om}{\om_b}\right)^{1/2} \left(\fr{\kappa}{0.5}\right){\rm erg}~{\rm cm}^{-3}~{\rm s}^{-1}, \label{loss1}
\end{split}
\eeq
with $\de (\om/\om_b)^{1/2} = (\om_2/\om_b)^{1/2} - (\om_1/\om_b)^{1/2}$, $\om_{\rm bi} \leq \om_1 < \om_2 \leq\om_b$ and $\om_1$ and $\om_2$ being the lower and upper limit of the emission band 
\citep[note also that the role of incoherent \u{C}erenkov radiation in the pressure balance within cool core clusters was addressed in][]{lie17}.

When the energy density of the hot ICM is divided by the volume emissivity of \eqref{loss1}, 
the ensuing time of emission could further be expressed as the number of wave oscillation cycles.   
The result for
the band $(\om_1= \om_{\rm bi},~\om_2= \om_b)$
ranges from $1$ to $10$~cycles across the band.  Thus, there is insufficient \u{C}erenkov radiation below the frequency $\om = \om_{\rm bi}$ to support even one cycle of oscillation,
and we shall ignore this
regime\footnote{Strictly speaking, one should use the second term on the right side of \eqref{nrni} instead of the first term to calculate the volume emissivity in the $\om < \om_{\rm bi}$ regime, but the result does not alter the conclusion that the radiation is emitted for less than one wave cycle; in fact, the problem becomes more severe the lower the frequency.},   This means only for frequencies $\om \gtrsim \om_{\rm bi}$ can the wave field accelerate charged particles to produce the non-thermal EUV radiation central to the theme of this paper.

Concerning the re-absorption of \u{C}erenkov radiation, 
the corresponding time scale, for $\om_{\rm bi} \leq \om \leq \om_b$, has been calculated in Section \ref{sec:coherent}, 
shown in equ. \eqref{tabs}.
For this $\om_{\rm bi} \leq \om\leq \om_b$ passband where there a significant amount of radiation and the emission timescale is of order a wave period or longer, but is much smaller than $\ta_{\rm abs}$, it should be realized that $\ta_{\rm abs}$ is nevertheless much smaller than the propagation time across the ICM.  Thus the \u{C}erenkov radiation occurs only for a short (\ie~$< \ta_{\rm abs}$)  time, during which the ICM exhibits fluctuations which perturb the medium from equilibrium (see Section \ref{sec:coherent}), but on scales greater than $\ta_{\rm abs}$ equilibrium is restored and the spectrum for the  $\om_{\rm bi} \leq \om\leq \om_b$ passband is Rayleigh-Jeans at temperature $kT \ap$~10 keV.

On timescales $< \ta_{\rm abs}$, however, the  $\om_{\rm bi} \leq \om\leq \om_b$ bandpass of 
\u{C}erenkov
radiation is capable of causing some ICM electrons to engage in a relativistic circular motion
\footnote{We ignored the influence of the frozen-in magnetic field of \eqref{cyclotron} because the wave field is a factor of $\gamma \gg 1$ larger.} 
if the charge has no initial velocity in the $\bk$ direction (any finite velocity along $\bk$ due to thermal motion is negligible in comparison to the transverse motion driven by the wave), with Lorentz factor $\gamma$ given (in the $\gamma \gg 1$ limit and as a function of the \u{C}erenkov radiation frequency) by 
\beq
\begin{split}
\gamma (\om) = & \fr{eE (\om)}{m_e c \omega}\\
= 
& 5.48 \left(\fr{n_e}{10^{-2}~{\rm cm}^{-3}}\right)^{1/2} 
\left(\fr{kT}{10~{\rm keV}}\right)^{1/2} \\
& \qquad\qquad\qquad\qquad\quad
\times\left(\fr{B}{1~\mu{\rm G}}\right)^{-1} \left(\fr{\om}{\om_b}\right)^{-3/4}
\label{gamma} 
\end{split}
\eeq 
(see \eg~equation (24.2) of \cite{sok86}, and equation (2) of \cite{kuo94}), where $E(\om)$ is the wave electric field,
and a fraction $\ap 10\%$ of the ICM energy is assumed to be spontaneously emitted as coherent \u{C}erenkov radiation.  
(Beyond the $10\%$ point, nonlinear feedback effects, due to significant changes in the magnitudes of the parameters of the hot gas,
greatly limits further \u{C}erenkov emission and invalidates the present calculation.)
This readily happens in an optically thin environment, as \eqref{loss1} indicates a significant fraction of the hot ICM energy is emitted on an $\ap 1$~s timescale; of course, the radiation is largely reabsorbed to replenish the ICM, although some of it is dissipated into particle acceleration -- the thrust of this paper.  The point is that until the \u{C}erenkov radiation energy is dissipated by an actual interaction, it remains part of the hot ICM energy budget, and there can be no observational ramifications.

Moreover, it should also be noted that there is a systematic reinforcement of the polarization of the 
\u{C}erenkov
radiation for varying $\hat\bk$.  Specifically, by summing $\bk \times \bj$ (see Section \ref{sec:coherent}) overall $\bk$ directions, the resulting vector has a large magnitude because the  $\mp$ circular polarization modes for the $\cos\th ^>_< 0$ hemispheres (see the discussion after (\ref{etasq_3})) always yield the {\it same} direction when $\bk \times \bj$ is projected onto the equatorial ($\th =0$) plane.  In fact, the lack of random cancellation of the polarization vectors across all \u{C}erenkov radiation propagation directions is because the refractive index selected only certain propagating modes, among which the polarization vectors are not randomly oriented.  Thus, from the discussions above, we may write 
\beq \fr{E^2}{4\pi} = U_\textrm{\u{C}erenkov} \ap U_{\rm ICM}.  \label{EU} 
\eeq

During the influence of \u{C}erenkov radiation,
the accelerated electron emits scattered radiation at the energy expense of the incident \u{C}erenkov radiation and at a synchrotron rate  
\beq 
\begin{split}
\fr{d\cE}{dt}\Bigg\vert_{\rm sync} = & \fr{2}{3} \fr{e^2 \om^2}{c} \gamma^4 \\
= & 1.44 \times 10^{-24} \left(\fr{n_e}{10^{-2}~{\rm cm}^{-3}}\right)^2 
\left(\fr{kT}{10~{\rm keV}}\right)^2 \\
& \qquad\qquad\quad \times
\left(\fr{B}{1~\mu{\rm G}}\right)^{-1} 
\left(\fr{\om}{\om_b}\right)^{-1}
~{\rm erg~s}^{-1}, \label{synch} 
\end{split}
\eeq
a formula derived by calculating the rate at frequency $\om$ and interval $\de\om = 0.1 \om$.
The typical outgoing (scattered) radiation has a frequency 
\beq 
\begin{split}
\om_{\rm sync} \ap \gamma^3 \om = 
& 2.88 \times 10^3 \left(\fr{n_e}{10^{-2}~{\rm cm}^{-3}}\right)^{3/2} 
\left(\fr{kT}{10~{\rm keV}}\right)^{3/2} \\
& \quad\times
\left(\fr{B}{1~\mu{\rm G}}\right)^{-3/4} \left(\fr{\om}{\om_b}\right)^{-5/4}
~{\rm rad~s}^{-1}. \label{sync} 
\end{split}
\eeq
The lifetime of the electron against the loss is 
\beq 
\begin{split}
\ta_{\rm sync} = & 3.12 \times 10^{18} 
\left(\fr{n_e}{10^{-2}~{\rm cm}^{-3}}\right)^{-3/2}  
\left(\fr{kT}{10~{\rm keV}}\right)^{-3/2} \\
& \qquad\qquad\qquad\qquad\times
\left(\fr{B}{1~\mu{\rm G}}\right)^{3/4} 
\left(\fr{\om}{\om_b}\right)^{1/4} ~{\rm s}. \label{tsync}
\end{split}
\eeq

The maximum number density of relativistic electrons having been accelerated to Lorentz factor $\gamma = \gamma (\om)$ of \eqref{gamma} by \u{C}erenkov radiation at frequency $\om$ and bandwidth $\de\om = 0.1 \om$  is given by the ratio of 
$U_\textrm{\u{C}erenkov} (\om) \de\om$ to $\gamma (\om) m_e c^2$, 
and is much less than the ICM number density $n_e$ even if the full range of $\gamma$ is considered.  
Specifically, this ratio is 
$
7.55\times 10^{-7}$~cm$^{-3}$ 
for the default parameter values used, and scales as $(n_e kT)^{1/2} B^{-1/4} \om^{5/4}$
if other values are used.  
Moreover, it is unchanged even if the charged particles are protons, although $\gamma(\om)$ is then given by \eqref{gamma} with $m_p$ replacing $m_e$.  
It should be emphasized, however, that until an interaction has taken place to enable a synchrotron photon or an inverse Compton photon (see below for the latter) to be emitted, no energy is removed from the ICM, as the \u{C}erenkov radiation is merely an intermediate state\footnote{The situation is analogous to Thomson scattering of an incident radiation beam by an electron. If during the time when the beam passes by the electron, no quantum interaction took place, which results in an outgoing photon, the electron would have just been accelerated by the oscillating electric field of the incident beam, but will remain at rest after the beam passing by (if it was initially at rest before the beam arrived) and no energy would have been removed from the beam.}, which will be reabsorbed.  
Thus, the fraction of ICM energy dissipated via \u{C}erenkov radiation depends on the frequency of interactions.

Apart from synchrotron radiation, the \u{C}erenkov radiation may also be dissipated via its acceleration of electrons to relativistic speed, enabling them to undergo IC scattering with the CMB to produce a net Poynting flux.  
The emission power per electron is $\si_{\rm T} \gamma^2 c U_{\rm CMB}$ where $\si_{\rm T}$ is the Thomson cross section, or 
\beq 
\begin{split}
\fr{d\cE}{dt}\Bigg\vert_{\rm IC} & = 5.00 \times 10^{-25} 
\left(\fr{n_e}{10^{-2}~{\rm cm}^{-3}}\right) 
\left(\fr{kT}{10~{\rm keV}}\right) \\
& \qquad\qquad\quad\times
\left(\fr{B}{1~\mu{\rm G}}\right)^{-1/2} 
\left(\fr{\om}{\om_b}\right)^{-3/2}
~{\rm erg~s}^{-1}, \label{IC} 
\end{split}
\eeq
The frequency of a scattered photon is typically 
\beq 
\begin{split}
\om_{\rm IC} = \gamma^2 \om_{\rm CMB} 
& = 3.04 \times 10^{13} \left(\fr{n_e}{10^{-2}~{\rm cm}^{-3}}\right) 
\left(\fr{kT}{10~{\rm keV}}\right) \\
& \quad\times\left(\fr{B}{1~\mu{\rm G}}\right)^{-1/2} 
\left(\fr{\om}{\om_b}\right)^{-3/2}
~{\rm rad~s}^{-1} \label{euv} 
\end{split}
\eeq
assuming the peak energy of a CMB photon to be $6.63 \times 10^{-4}$~eV.  
The lifetime against IC losses is 
\beq 
\begin{split}
\ta_{\rm IC} & = 1.79 \times 10^{19} \left(\fr{n_e}{10^{-2}~{\rm cm}^{-3}}\right)^{-1/2} \left(\fr{kT}{10~{\rm keV}}\right)^{-1/2} \\
& \qquad\qquad\qquad\qquad\times
\left(\fr{B}{1~\mu{\rm G}}\right)^{1/4} 
\left(\fr{\om}{\om_b}\right)^{3/4}
~{\rm s}. \label{taIC} 
\end{split}
\eeq  
We now explore the consequences of the two dissipation mechanisms of this section and the absorption process of the last.


According to the previous two sections, there are three timescales at play: the absorption time $\ta_{\rm abs}$ of \eqref{tabs}, the synchrotron loss time $\ta_{\rm sync}$ of \eqref{tsync}, and the IC loss time $\ta_{\rm IC}$ of \eqref{IC}.  
The first transfers the \u{C}erenkov radiation energy (spontaneous and stimulated) back to the hot ICM, while the second and third are the genuine loss mechanisms which remove energy from the \u{C}erenkov emission (hence the ICM electrons) by converting it to another form ahead of absorption.  
By comparing \eqref{tsync} to \eqref{IC}, however, it is clear that for most frequencies $\om < \om_b$ of interest $\ta_{\rm sync} > \ta_{\rm IC}$, \ie~synchrotron losses occur more slowly than IC losses, and we shall henceforth ignore the former.

\section{Spectrum of Non-thermal Emission}
\label{sec:entropy}
The spectrum of the radiation is, according to \eqref{differential}, $\propto \om^{-1/2}$ for $\om \leq \om_b$.  Consider a narrow bandwidth $\de\om$ at $\om$, 
and write its share of the total emitted ICM energy as \beq p(\om) \de\om = \fr{1}{2\sqrt{\om_b \om}} \de\om \label{pw} \eeq  Since for the frequency range of interest, $\om_{\rm bi} \leq \om \leq \om_b$,  the 
\u{C}erenkov absorption time  of (\ref{tabs}) satisfies the inequality $\ta_{\rm abs} < \ta_{\rm IC}$, the quotient $x = \ta_{\rm abs}/\ta_{\rm IC}$ will give the fraction of \u{C}erenkov energy dissipated before reabsorption.  Now the time it takes for continuous absorption within this narrow band to remove the equivalent of {\it all} the energy of the \u{C}erenkov radiation  is $y = \ta_{\rm abs}/(p(\om)\de\om)$, so the time to dissipate this equivalent energy by IC emission is evidently $z=y/x = \ta_{\rm IC}/(p(\om)\de\om)$.  Dividing the Hubble time by $z$ then yields the fraction of total \u{C}erenkov energy dissipated through this bandwidth during the lifetime of a cluster.
Lastly, one may integrate from $\om =\om_{\rm bi}$ to $\om=\om_b$ with $p(\om)$ as defined in (\ref{pw}).  
The result is
\beq
\begin{split}
\fr{U_{\rm IC}}{U_{\rm ICM}}
& = 
0.0214
\left(\fr{n_e}{10^{-2}~{\rm cm}^{-3}}\right)^{-1/6} 
\left(\fr{kT}{10~{\rm keV}}\right)^{-1/6} \\
& \qquad\qquad\qquad\qquad\times
\left(\fr{B}{1~\mu{\rm G}}\right)^{-1/6}  \left(\fr{h}{0.7}\right)^{-1}
\label{totaloss}    
\end{split} 
\eeq
where $h$ is the Hubble constant in units of 100~km~s$^{-1}$~Mpc$^{-1}$.

Thus 
$\ap98\%$
of the hot ICM survives the IC emission over a cluster's lifetime, with the relativistic electrons being available because of acceleration by \u{C}erenkov radiation.  Since this radiation is negligible at frequencies $\om < \om_{\rm bi}$, \eqref{euv} shows that the maximum frequency of the IC scattered radiation is in the X-ray range, corresponding to $\om \ap \om_{\rm bi}$, where $\ta_{\rm IC}$ falls below a Hubble time.  Such X-rays would have already been entirely emitted by now.

As shown below, the lowest frequency, at which $\ta_{\rm IC}$ 
barely
falls beneath a Hubble time, is given by \eqref{wcritwb} and is $\ll\om_b$.  
Integrating $U_{\rm IC}$ in \eqref{totaloss}, with $U_{\rm ICM}$  given by \eqref{cE}, from this frequency upwards to $\om_b$, and multiplying the result by the ratio of the cluster radius $R$ to the Hubble radius (because the IC scattered photons comprising the energy density $U_{\rm IC}$ are in an optically thin ICM and, as such, have constantly been freely streaming out of the cluster), yields a total luminosity from the infrared 
frequency upwards, of
\beq
\begin{split}
L_{\rm cluster} & = 
4.64\times10^{43}
\left(\fr{n_e}{10^{-2}~{\rm cm}^{-3}}\right)^{5/6}
\left(\fr{kT}{10~{\rm keV}}\right)^{5/6} \\
& \qquad\qquad\times
\left(\fr{B}{1~\mu{\rm G}}\right)^{-1/6}  \left(\fr{R}{1~{\rm Mpc}}\right)^3
~{\rm ergs~s}^{-1}. \label{IRFUV} 
\end{split}
\eeq
This value is 
three
orders of magnitude beneath the observed optical luminosity of the Coma cluster
\citep{fus94}.

It should also be remarked that the radiation emerging from a direct non-linear scattering between the ICM electrons and the \u{C}erenkov radiation has, by \eqref{sync}, a frequency in the MHz range or lower, \ie~such radio signals do not pass through the ionosphere and cannot be observed.

The remaining task is to evaluate the EUV brightness of a cluster, which requires special attention because for all \u{C}erenkov incoming frequencies $\om < \om _{\rm crit}$, the IC losses would reduce the Lorentz 
factor
of the relativistic electrons to the EUV emitting range of $\gamma \ap 300$ well within one Hubble time, so that relativistic electrons riding \u{C}erenkov amplitudes of the $\om < \om _{\rm crit}$ modes could be easily lost the bulk of their energies to reach $\gamma\ap 300$ in a cluster's lifetime. Once they are at this Lorentz factor or thereabout, they will emit EUV radiation for an entire Hubble time (or a cluster's lifetime).  Thus one must calculate the EUV band more carefully.

Since the Lorentz factor of the electrons is $\gamma (\om)$ as given by \eqref{gamma}, the fraction of total 
\u{C}erenkov,
or 
$10\%$ of the
hot ICM, energy residing in this narrow band as $\gamma \ap 300$ electrons responsible for the EUV excess emission via IC emission 
\citep[see also \eqref{euv}]{sar98} is 
\beq 
\begin{split}
\fr{U(\om, \gamma = 300)}{U_{\rm ICM}} & 
= \fr{1}{H_0 z} \times \fr{300}{\gamma (\om)} \times 
0.1 \\
& = \fr{
30\,
U_{\rm CMB} \si_{\rm T} p(\om)\de\om}{H_0 m_e c}, \label{Uomg} 
\end{split}
\eeq 
where it is understood \citep{sar98} that $\gamma \ap 300$ electrons can just about survive a Hubble time against IC interaction with the CMB.  More precisely, \citet{sar98} showed that $\gamma > 300$ electrons could undergo IC emission (at soft X-ray or higher energies) to reach $\gamma \ap 300$ within the lifetime of a cluster.

Energetic components in the galactic cluster have well-established, such as relativistic protons in cosmic ray, which is responsible for Gamma-ray emission, observed in Coma cluster by Fermi-Lat \citep{Adam2021}, in their hadronic collision with ICM, and relativistic electrons with a Lorentz factor $>1000$. 
These electrons produce the so-called diffuse radio emissions, which are not directly associated with any radio galaxy and are detected in deep radio observations \citep[i.e.][]{Ferrari2008,Bykov2019,vanWeeren2019} in the last decade.
Shock waves in the scales from 10 kiloparsecs up to megaparsecs, arising in the processes of accretion or merging, are considered to be a promising mechanism for accelerating particles in the presence of a magnetic field.
Some of these shock waves are associated with a class of diffuse radio sources with strong polarizations, referred to as "radio relics" in deep radio observations. 
X-ray emissions are also detected, associated with large-scale shock waves in ICM \citep{chu22}.
It is widely adopted that, in the presence of an inhomogeneous magnetic field, shocks transfer a part of the energy of plasma motions in ICM to energetic particles in the manner called the "first-order Fermi acceleration". 
In this mechanism, particles run across the front of the shock wave backwards and forward as when being scattered by the inhomogeneous magnetic field. 
During those scatters, articles gain some energy each time crossing the front of a shock wave. 
The acceleration is also made by the random and inhomogeneous fluctuations in the magnetic field triggered by turbulence. 
This is the second-order Fermi acceleration.
But this mechanism is not expected to be as efficient as the previous one because of the randomness it involves \citep{Bykov2019,vanWeeren2019}.
Nevertheless, the relativistic electrons that we discuss here are mainly accelerated by the primary mechanism, so these cannot be pumped up to GeV. 
We'll skip the acceleration mechanism of shock waves.

The total energy density of the EUV emitting electrons is obtained by integrating \eqref{Uomg} over all frequencies to $\om_{\rm crit}$, the value of $\om$ at which $\gamma (\om) \to 300$ from above (see (\ref{gamma})) as the cluster ages.  There is no need to integrate above $\om_{\rm crit}$ because as $\om$ exceeds $\om_{\rm crit}$,  $\gamma (\om)$ drops below $300$, and it is impossible for IC losses to access $\gamma\ap 300$ from a $\gamma < 300$ regime.  From \eqref{gamma}, one may solve for $\om=\om_{\rm crit}$, namely 
\beq 
\begin{split}
\fr{\om_{\rm crit}}{\om_b} = 
2.23 & \times 10^{-3} 
\left(\fr{n_e}{10^{-2}~{\rm cm}^{-3}}\right)^{2/3} \\
& \times\left(\fr{kT}{10~{\rm keV}}\right)^{2/3} \left(\fr{B}{1~\mu{\rm G}}\right)^{-1/3}
~{\rm rad~s}^{-1} \label{wcritwb} 
\end{split}
\eeq  
The above integration may then be performed to yield 
\beq 
\begin{split}
r_{\rm EUV} = \fr{U_{\rm EUV}^{\rm e}}{U_{\rm ICM}} 
& = \fr{
30 U_{\rm CMB} \si_{\rm T}}{H_0 m_e c} \int_0^{\om_{\rm crit}} p(\om) d\om \\
& = 
1.79 \times 10^{-3} 
\left(\fr{n_e}{10^{-2}~{\rm cm}^{-3}}\right)^{1/3}   \\
& \times
\left(\fr{kT}{10~{\rm keV}}\right)^{1/3}
\left(\fr{B}{1~\mu{\rm G}}\right)^{-1/6} 
\left(\fr{h}{0.7}\right)^{-1}.
\label{euvr} 
\end{split}
\eeq
The pressure ratio of the EUV emitting relativistic electrons to the hot ICM is $r_{\rm EUV}/2$, and for default parameters
pars with 
the estimate of \citet{sar98} as the pressure required of cosmic rays 
(
were they the source of the clusters EUV excess emission via IC scattering of the CMB
).

In fact, one can divide $U_{\rm EUV}^e$ by the emission time $\ta_{\rm IC}$ of (\ref{IC}), 
with $\om$ in the latter {\it fixed} at $\om=\om_{\rm crit}$ of (\ref{wcritwb}), 
to obtain the EUV volume emissivity $j_{\rm EUV}$.  
Of course, use has to be made of the relation 
\beq
\begin{split}
U_{\rm ICM} &= \fr{\cE}{V} = 3n_e kT \\
& = 4.8 \times 10^{-11} \left(\fr{n_e}{10^{-3}~{\rm cm}^{-3}}\right) 
\left(\fr{kT}{10~{\rm keV}}\right)~{\rm erg}~{\rm cm}^{-3} \label{cE} 
\end{split}
\eeq 
to arrive at the answer, 
which may then be multiplied by the volume $V=4\pi R^3/3$ of the entire cluster to obtain the total EUV brightness 
\beq 
\begin{split}
L_{\rm EUV} & =
3.79 \times 10^{44} 
\left(\fr{n_e}{10^{-2}~{\rm cm}^{-3}}\right)^{4/3}
\left(\fr{kT}{10~{\rm keV}}\right)^{4/3}\\ 
& ~\times 
\left(\fr{B}{1~\mu{\rm G}}\right)^{-1/6} \left(\fr{R}{1~{\rm Mpc}} \right)^3 \left(\fr{h}{0.7}\right)^{-1}
~{\rm ergs~s}^{-1}. \label{lumin}    
\end{split} \eeq  
According to observations by 
\citet{
bon03}, such a brightness is on par with the observed 74 -- 203 eV luminosity of A1795 in the 3' -- 10', or $(0.21 -- 0.72) h_{0.7}^{-1}$ Mpc, region; and Coma cluster 0' -- 90', or $2.6h_{0.7}^{-1}$ Mpc radius.  If the EUV excess emission from these radii is of a thermal origin, however, the mass of the $\ap 10^6$~K warm gas will be colossal, being in the $10^{14-15}~M_\odot$ range as noted by 
\citet{
bon03}.  
The purpose of this paper is to show that the non-thermal interpretation of the EUV excess as IC scattering between the CMB and relativistic ICM electrons, accelerated by the low frequency and large amplitude fields of \u{C}erenkov radiation, 
is a viable alternative, especially for the inner radii of a cluster.

\section{Summary and conclusion}

It is shown that the key and representative parameters of the magnetoionic medium of a cluster of galaxies, namely $kT \ap 10$~keV, $n_e \ap 10^{-3}$~cm$^{-3}$, and $B \ap 1~\mu$G coherent over $\ap 1$~Mpc length scale, indicates the medium supports coherent \u{C}erenkov emission capable of propagating at frequencies $\om$ below the ICM plasma frequency  $\om_p$.  The radiation has an intensity spectrum proportional to $\om^{1/2}$, between $\om=\om_{\rm bi}$ and $\om=\om_b$ where $\om_b$ is the electron cyclotron frequency.  
For $\om > \om_b$, the radiation can no longer propagate through the ICM plasma, and the spectrum cuts off.  On the other hand, for $\om < \om_{\rm bi}$ there is insufficient 
\u{C}erenkov
radiation to last one cycle of wave oscillation, and the effect can be ignored.

The significance of the low-frequency 
\u{C}erenkov
radiation is in the largeness of the $E/\om$ ratio where $E$ is the r.m.s. wave electric field, which enables the wave to accelerate charged particles to relativistic speeds along a circular orbit.  Such electrons will, in turn, emit synchrotron radiation and undergo IC scattering with the CMB.
The outgoing (scattered) radiation
is in the EUV range of frequencies, and gives rise to a total EUV luminosity $\ap 10^{44}$~ergs~s$^{-1}$ for the entire cluster, which is in broad agreement with observations.

Thus the model developed here substantiates the earlier ideas of \citet{hwa97,ens98,sar98}, who proposed IC scattering as the non-thermal origin of the EUV excess of clusters discovered in the mid-90s and revisited recently by e-Rosita.  It should be mentioned that he results presented were derived using a quasi-linear approach to the problem to highlight the salient and observable features of the phenomenon, namely the EUV emission, even though the large amplitude coherent \u{C}erenkov modes which source the emission are nonlinear (see the portion of section 3 immediately after (\ref{gamma})).  An in-depth analysis of e-Rosita data in the near future will shed further light on the question whether a cluster's EUV excess is due to a thermal or non-thermal process.

\section*{Acknowledgements}
This work is sponsored (in part) by the Chinese Academy of Sciences (CAS), through a grant to the CAS South America Center for Astronomy (CASSACA) in Santiago, Chile. 

\section{Data Availability}
No new data were generated or analysed in support of this research.



\bibliographystyle{mnras}

\appendix
\section{The Refractive Index in a Hydrogen Plasma}
\label{appendix::RefractiveIndex}

In a hydrogen plasma, when the contribution from ions is considered,
it is convenient to introduce two dimensionless parameters 
$\xi_i\equiv\omega_{\rm pi}^2/\omega^2$ and $\eta_i\equiv\omega_{\rm bi}^2/\omega^2$ analogous to \eqref{xi_eta},
where $\omega_{pi} =\sqrt{n_ee^2/(\epsilon_0 m_p)}$ is the plasma frequency corresponding to protons with a mass of $m_p$ and $\omega_{bi}=|e|B/m_p$ is the proton cyclotron frequency. 
It is clear that $\omega_{\rm pi}$ and $\omega_{\rm bi}$ are much lower in frequencies than $\omega_p$ and $\omega_b$, respectively, namely,
\begin{equation}
\omega_{\rm bi}/\omega_{b} = (\omega_{\rm pi}/\omega_p)^2 
= m_e/m_p.
\label{OmegaRate} 
\end{equation}
When radiation propagates along the direction of an angle $\theta$ away from the magnetic field, ions contribute the refractive index $n$ of the medium in the form of 
\begin{equation}
-\xi_{i}\bigg[1 + \frac{\eta_i\sin^2\theta}{2\xi_i} 
\pm \Big(\frac{\eta_i^2\sin^4\theta}{4\xi_i^2} + \eta_i\cos^2\theta\Big)\bigg].
\label{IonRefractionIndex}
\end{equation}
This result can be yielded by simplifying \citet{jin06}'s work in the regime $\xi_i\gg\eta_i$ and $\xi_i\gg1$ when following the similar derivation of \eqref{etasq}.

The terms of $\cos^2\theta$ in both \eqref{etasq} and \eqref{IonRefractionIndex} dominate the $\sin^2\theta$ and $\sin^4\theta$ apart from the angle interval $\delta\theta$, evaluated in \eqref{cone}, on either side of $\theta=\pi/2$.
We set $\cos\theta=1$ to reveal the salient feature of radiation propagation in ICM. 
\eqref{etasq} and \eqref{IonRefractionIndex} can be reduced into \eqref{etasq_3}, and a step further into
\begin{equation}
\begin{split}
n^2=  1-\fr{\om_p^2/\om^2}{1 + \om_b/\om} -
& \fr{\om_{\rm pi}^2/\om^2}{1 - \om_{\rm bi}/\om};\\
& \qquad
 \om \ll \om_{\rm pi}\ll\om_p~{\rm and}~\om_b \ll \om_p. \label{etasq_4}
 \end{split}
\end{equation}
For radiation frequency satisfies the above condition, it is trivial to calculate that the leading term of $n^2$ is $\omega_p^2/(\omega\omega_b)$ when $\omega_{\rm bi}<\omega <\omega_b$. 
While the radiation frequency drops further to lower than the proton cyclotron frequency, viz. $\omega<\omega_{\rm bi}$, we can write the square of the refractive index as
\begin{equation}
\begin{split}
n^2 & = 1 
  - \frac{\omega_p^2}{\omega\omega_b}\bigg(1 + \frac{\omega}{\omega_b}\bigg)^{-1} 
  + \frac{\omega_{\rm pi}^2}{\omega\omega_{\rm bi}}\bigg(1 - \frac{\omega}{\omega_{\rm bi}}\bigg)^{-1}\\
  &\approx 1 - \frac{\omega_p^2}{\omega\omega_b} 
  + \frac{\omega_p^2}{\omega_b^2} 
  + \frac{\omega_{\rm pi}^2}{\omega\omega_{\rm bi}} 
  +  \frac{\omega_{\rm pi}^2}{\omega_{\rm bi}^2}
\end{split}
\end{equation}
Substituting \eqref{OmegaRate} into the last line of the above calculation, 
the second and fourth terms cancel each other. 
Then there are only three terms left, namely
\begin{equation}
n^2 = 1 + \frac{\omega_p^2}{\omega_b^2} 
  +  \frac{\omega_p^2}{\omega_{\rm bi}\omega_b}.
\end{equation}
Thus the last term is the leading one due to the fact that $\omega_b\gg\omega_{\rm bi}$.
When combining our calculations in two cases that $\omega$ varying across $\omega_{\rm bi}$, the refractive index $n$ is shown as 
\begin{equation}
n \approx \frac{\omega_p}{\sqrt{\omega\omega_b}}\Theta(\omega-\omega_{\rm bi})
+ \frac{\omega_p}{\sqrt{\omega_{\rm bi}\omega_b}}\Theta(\omega_{\rm bi}-\omega)
\end{equation}
with the aid of the Heaviside unit step function $\Theta(\omega)$.

\numberwithin{equation}{section}

\bsp	
\label{lastpage}
\end{document}